\documentclass[%
twocolumn,
superscriptaddress,
 amsmath,amssymb,
 aps,
prc,
floatfix,
]{revtex4-1}

\usepackage{amsmath}
\usepackage{amssymb}
\usepackage{amsfonts}

\usepackage{xcolor}

\usepackage{dsfont}

\usepackage{braket}
\usepackage{graphicx}
\usepackage{dcolumn}
\usepackage{bm}

\begin{document}
\title{Clusterization in nuclear states at the edge of stability }
%
%

\author{J.P. Linares Fernandez}
\affiliation{Department of Physics and Astronomy, Louisiana State University, Baton Rouge, LA 70803, USA}
\email{jlinares@lsu.edu}
\author{N. Michel}
\affiliation{AS Key Laboratory of High Precision Nuclear Spectroscopy, Institute of Modern Physics, Chinese Academy of Sciences, Lanzhou 730000, China}
\email{nicolas.lj.michel@outlook.fr}
\affiliation{School of Nuclear Science and Technology, University of Chinese Academy of Sciences, Beijing 100049, China}
\author{M. P{\l}oszajczak}
\affiliation{Grand Acc\'el\'erateur National d'Ions Lourds (GANIL), CEA/DSM - CNRS/IN2P3, BP 55027, F-14000 Caen, France}
\email{marek.ploszajczak@ganil.fr}

\begin{abstract}
    The open quantum system eigenstate in the vicinity of low-energy decay channel may mimic its features, in particular the characteristic clustering properties of the decay  channel. This generic mechanism of clusterization, the so-called mimicry mechanism of clusterization,  is discussed here on example of the ground state wave function of $^8$Be. At higher excitation energies, when the density of states and reaction channels is high, the quantal aspects in the clusterization process disappear and the statistical mechanism of clusterization which is rooted in the Central Limit Theorem, begin to dominate.
\end{abstract}
\maketitle
\section{Introduction}
\label{intro}
Clustering is ubiquitous in Nature and clearly one of the most mysterious processes in Physics. It happens at all 
scales in time, distances and energies: from the microscopic scales of hadrons and nuclei to the macroscopic 
scales of living organisms and clusters of galaxies, from the high excitation energies to cold systems.
There are many specific reasons for the cluster production but there are also few generic mechanisms of 
the clusterization which are independent of individual features of the studied system. 

In nuclear physics, there are two such generic mechanisms. At high excitation energies, the classical aspects prevail and it is the statistical mechanism rooted in the Central Limit Theorem which describes the clusterization \cite{botet_book_2002}. At low excitation energies, close to the threshold of reaction channel, clustering appears as the emergent phenomenon due to the interaction between the system and its environment of decay channels and scattering continuum. 

After a short discussion of the statistical mechanism, we will concentrate on the presentation of mimicry mechanism of clusterization and discuss few examples in $^8$Be.

\section{Statistical mechanism of clusterization}
\label{stat}
At excitation energies around Fermi energy or higher, the quantum features in the cluster production are largely unimportant. In this classical regime, one can identify two distinct statistical scenarios: the fragmentation scenario described by various hybrids of the Fragmentation-Inactivation Binary (FIB) model (see Ref. \cite{botet_book_2002} and references cited therein), and the aggregation scenario which includes a broad family of both equilibrium models, like Fisher droplet model, Ising model, or percolation model, and non-equilibrium models, like the Smoluchowski model of gelation \cite{botet_book_2002}. 

In the cascading process of the FIB, whole information is contained in the fragmentation kernel ${\cal F}_{j,k-j}$ which tells what is the probability that a cluster of a size $k$ fragments into fragments $j$ and $k-j$, and the inactivation kernel ${\cal I}_k$ which prescribes the probability that cluster a $k$ remains inactive. If the fragmentation happened then the fragment $k-j$ can split into smaller fragments or remain inactive. The cascade stops when all fragments become inactive. The details of the FIB model can be found in Ref. \cite{botet_book_2002}.
A typical example of the aggregation scenario is the Smoluchowski model where the time-evolution of the cluster population is governed by the aggregation kernel ${\cal A}_{i,j}$ which tells the fusion probability of clusters $i$ and $j$.

The basic tool to study various phases of the fragmentation or aggregation processes is the $\Delta$-scaling \cite{PhysRevE.62.1825}:
\begin{eqnarray}
\langle m \rangle^{\Delta} {\cal P}_{\langle m \rangle}[m] &=& \Phi \left( z_{(\Delta)}\right) ~~~~~~~~~~~0 <\Delta\leq 1     \nonumber  \\
z_{(\Delta)} &=& (m-m^*)/\langle m \rangle^{\Delta}
\end{eqnarray}
where ${\cal P}_{\langle m \rangle}[m]$ is the normalized probability of the variable $m$ for different 'system sizes' $<m>$, and $m^*$ is the most probable value. If the scaling holds then the scaling relation holds independently of any phenomenological reasons to change $<m>$. In the ordered phase, the system exhibits the so-called second scaling law with $\Delta = 1/2$, whereas in the disordered phase first scaling with $\Delta = 1$ holds. 

These two generic clusterization scenarios are characterized by the different order parameter. In the fragmentation scenario, order parameter is the average cluster multiplicity, whereas in the aggregation scenario it is the average size of the largest fragment. Also the cluster size distribution is different in both scenarios. At the critical point of the fragmentation scenario, the cluster-size distribution is $n(s) \sim s^{\omega}$ with $\omega < 2$, whereas in the aggregation scenario the exponent of the power-law distribution is $\omega > 2$. 

The experimental analysis of the cluster production in central heavy-ion (HI) collisions with $E_{\rm lab} \geq 25$ MeV showed that the order parameter in the clusterization process  is the average size of the largest fragment and, hence, the fragment production obeys  the aggregation scenario \cite{PhysRevLett.86.3514}. More about the clusterization and fragment production in HI collisions at energies above the Fermi energy can be found in Ref. \cite{botet_book_2002}.

\section{Shell model for open quantum systems}
\label{gsm}
An open quantum system is a quantum system which is found to be in interaction 
with an external quantum system, the environment. The open quantum system can 
be viewed as a distinguished part of a larger quantum system which is usual considered as a close quantum system (CQS). Standard techniques developed in the context of open quantum systems are based on the determination of the density matrix for the subsystem: $\rho_{\rm S} = {\rm Tr}_{\rm E}\left\{\rho_{\rm full}\right\}$, where the trace is taken over states of the environment. This technique have proven powerful in fields such as: quantum optics, quantum measurement theory, quantum statistical mechanics, quantum information science, quantum cosmology, mesoscopic physics, and others. 
Description in terms of the density matrix is not well suited for nuclear physics which deals with the well-defined quantum states for which shell model formulation is advantageous.

Near-threshold states cannot be described in the closed quantum system framework which violates unitarity at the particle emission threshold. An appropriate theoretical tool to study the clusterizatrion in these states is the shell model (SM) for open quantum systems (OQSs). The first attempt in this direction was the continuum shell model in Hilbert space \cite{Mahaux_Weidenmuller, Barz78, Phi77, Rotter78, Okolowicz2003, Bennaceur99, Bennaceur00, ROTUREAU2006}, the so-called shell model embedded in the continuum, which  provides a unified description of structure and reactions.

In this work, we will concentrate on the discussion of SM for OQSs in Berggren basis, the so-called Gamow shell model (GSM) \cite{michel_review_2009,Michel2002,PhysRevLett.89.042501,PhysRevC.70.064313,michel_book_2021}, which is formulated in two different representations. 
The Slater determinant representation of the GSM provides a natural generalization of the standard SM. As its predecessor, the GSM is the theoretical tool for nuclear structure studies as the asymptotics of individual reaction channels cannot be correctly defined. The second representation is formulated in the coupled reaction channels (GSM-CC) \cite{Fossez15,mercenne_2019,PhysRevC.89.034624,michel_book_2021}. In this representation both spectra and reaction cross sections can be defined in the unifying framework. 

In GSM-CC, the reaction channels are defined as:
\begin{equation}
|c,r> = {\cal A}\left( |\Psi_{\rm T}^{J_{\rm T}}\rangle \otimes |\Psi_{\rm P}^{J_{\rm P}}\rangle \right)_{M_{\rm A}}^{J_{\rm A}} \ ,
\end{equation}
where the indices T and P refer to target and projectile, respectively. $\Psi_{\rm T}$ and $\Psi_{\rm P}$ are GSM wave functions and ${\cal A}$ is the antisymmetrizer. The quantum number $c \equiv \left( Z - z, N - n, J_{\rm T} ; z, n, \ell, J_{\rm int}, J_{\rm P} \right)$ includes the quantum numbers of both target and projectile which are proton number, neutron number, and spin. $\ell$ is the angular momentum of the projectile, and $J_{\rm int}$ is the intrinsic spin of the projectile. The spins are coupled as ${\bf \ell} + {\bf J_{\rm int}} = {\bf J_{\rm P}}$ and ${\bf J_{\rm T}} + {\bf J_{\rm P}} = {\bf J_{\rm A}}$. The channel completeness relation:
\begin{equation}
\sum_c\int_0^{\infty} dr r^2 |c,r\rangle \langle c,r| = {\bf 1}
\label{eq2}
\end{equation}
allows to expand any solution of the Hamiltonian:
\begin{equation}
|\Psi\rangle_{M_{\rm A}}^{J_{\rm A}} = \sum_c\int_0^{\infty} dr r^2\frac{u_c^{J_{\rm A}}(r)}{r} |c,r\rangle \ ,
\label{eq1}
\end{equation}
where $u_c^{J_{\rm A}}$ is the radial amplitude describing the relative motion of the projectile with respect to the core with a total angular momentum $J_{\rm A}$. Inserting (\ref{eq1}) into the Schr\"odinger equation one derives the coupled-channel equations in a form similar to those of the resonating group method. Details of the theoretiical framework and the methods to solve the GSM-CC equations can be found in Ref. \cite{michel_book_2021}.

\subsection{Spectroscopy of $^8$Be}
\label{gsm_appl}
$^8$Be nucleus plays an important role in the He-burning process, particularly in the formation of the Hoyle resonance $0_2^+$ in $^{12}$C, which ultimately leads to the production of $^{16}$O. The description of low-energy states in $^8$Be demands the inclusion of the coupling to $^4$He continuum. In addition, we use the following  mass partitions: $^7$Be + n, $^7$Li + p, and $^6$Li + d in the completeness relation (\ref{eq2}).

\subsubsection{Model space}
The GSM calculations of $^{6,7}$Li and $^{7,8}$Be are done with a $^4$He core and two or three valence nucleons (one proton and one neutron for $^6$Li, and two protons(neutrons) and one neutron(proton) for $^7$Be($^7$Li)). Hence, the $0s_{1/2}$ harmonic oscillator (HO) shells are fully occupied and inert. The valence space consists of two resonant-like HO shells $0p_{3/2}, 0p_{1/2}$ and several scattering-like subdominant HO shells in $\{s_{1/2}\}$, $\{p_{3/2}\}$, $\{p_{1/2}\}$, $\{d_{5/2}\}$, $\{d_{3/2}\}$, $\{f_{7/2}\}$, $\{f_{5/2}\}$, $\{g_{9/2}\}$, and $\{g_{7/2}\}$ partial waves, with $n$ in the interval $[$1 : 4$]$ for single particle waves and $[$0 : 4$]$ in other partial waves. The small number of scattering-like HO states approximate the non-resonant continuum, and additionally, this approximation reduces the size of the GSM matrix. 

The basis of Slater determinants is truncated by limiting the excitation energy to 8$\hbar\omega$. The internal structure of $^4$He-projectiles is calculated using the N$^3$LO interaction without a three-body contribution. The N$^3$LO realistic interaction is diagonalized in eight HO shells to generate intrinsic states of $^4$He. The oscillator length chosen for this calculations is $b$ = 1.7 fm. With this space and oscillator length, the calculated energy of the ground state $0^+_1$ of $^4$He is -26.32 MeV, whereas the experimental value is -28.30 MeV. Using the same parameters and interaction, the calculated energy of the deuteron projectile is -1.95 MeV, whereas the experimental value is -2.2 MeV. In the calculations, we use the experimental binding energy of $^4$He to assure the correct energy of the $^4$He + $^4$He threshold.

Channels with one-proton (one-neutron) projectile are built by coupling the $^7$Li ($^7$Be) wave functions having $K_i^{\pi}$ = $3/2_1^-$, 
$1/2_1^-$, $7/2_1^-$, $5/2_1^-$, $5/2_2^-$, $3/2_2^-$, $1/2_2^-$, $7/2_1^-$ with the proton (neutron) wave functions in the partial waves 
${\ell j:} s_{1/2}, p_{1/2}, p_{3/2}, d_{3/2}, d_{5/2}, f_{5/2}, f_{7/2}, g_{7/2}, g_{9/2}$. The cluster channels 
[$^4$He(${0_1^+} \otimes {^4{\rm He}}(L_{\rm CM} J_{\rm int} J_{\rm P})]^{J^{\pi}}$ are constructed by coupling $^4$He wave function in partial waves: 
${^1S_0}, {^1P_1}, {^1D_2}, {^1F_3}, {^1G_4}$ with the $^4$He core in $K_i^{\pi} = 0_1^+$ state. 
The deuteron channels $[{^6{\rm Li}}(K_i^{\pi}) \otimes {^2{\rm H}}(L_{\rm CM} J_{\rm int} J_{\rm P})]^{J^{\pi}}$ are constructed by coupling the deuteron wave function in partial waves: ${^3S_1}, {^3P_0}, {^3P_1}, {^3P_2}, {^3D_1}, {^3D_2}, {^3D_3}$, with the $^6$Li target in $K_i^{\pi} = 1_1^+, 3_1^+, 0_1^+, 2_1^+, 2_2^+, 1_2^+$ states.

\subsubsection{Spectrum of $^8$Be}
\begin{figure}[h]
\centering
\includegraphics[width=7cm,clip]{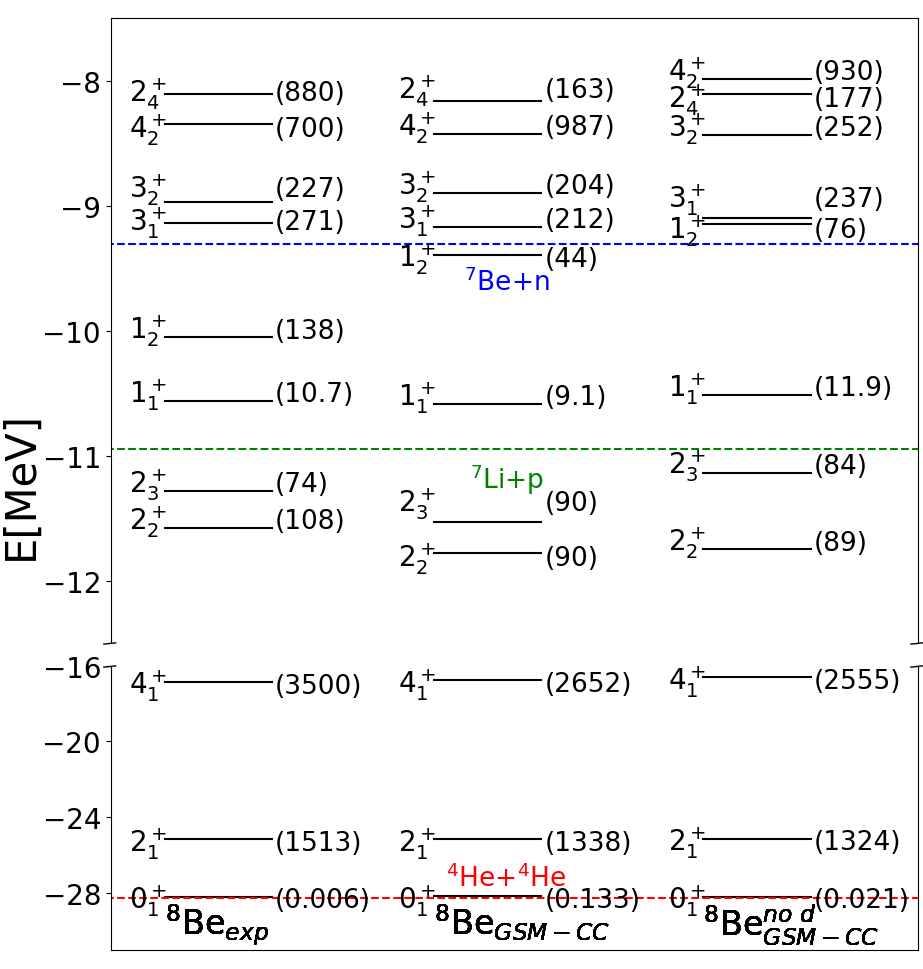}
\caption{Energy spectrum of $^8$Be calculated in GSM-CC is compared with the experimental spectrum 
and a spectrum calculated without the deuteron channels. Numbers in the brackets indicate the resonance widths in keV. All energies are given relative to the energy of $^4$He core. Experimental particle emission thresholds are shown with dashed lines.}
\label{fig-0}      
\end{figure}

The spectrum of $^8$Be is calculated in the channel basis comprising the [$^4$He($0^+_1) \otimes ^4$He($L_{\rm CM} J_{\rm int} J_{\rm P}$])]$^{J_{\pi}}$ , [$^6$Li($K_i^{\pi}$)$\otimes ^2$H($L_{\rm CM} J_{\rm int} J_{\rm P}$])]$^{J_{\pi}}$, [$^7$Li($K_i^{\pi}) \otimes$ p($\ell j)]^{J_{\pi}}$, and [$^7$Be($K_i^{\pi}) \otimes$ n($\ell j)]^{J_{\pi}}$ channels. Figure \ref{fig-0} shows the spectrum of $^8$Be calculated with and without the deuteron channel. The agreement with the experimental data for resonance energies and widths is good, especially if the deuteron channels are included. The inclusion of deuteron channels  [$^6$Li($K_i^{\pi}$)$\otimes ^2$H($L_{\rm CM} J_{\rm int} J_{\rm P}$])]$^{J_{\pi}}$ helps to close the gaps between states of each doublet: ($1_1^+; 1_2^+$), ($2_1^+; 2_2^+$), ($3_1^+; 3_2^+$). 
For $2_1^+$, $4_1^+$ resonances, the probability of deuteron reaction channels surpass even the probability of $\alpha$-channels: [$^4$He$(0_1^+) \otimes ^4$He$(L_{\rm CM} J_{\rm int} J_{\rm P})]^{J^{\pi}}$.

\section{Mimicry mechanism of clusterization}
Almost 60 years ago, Ikeda et al  have noticed that $\alpha$ cluster states can be found in the proximity of $\alpha$ decay threshold \cite{1968PThPS..68E.464I}. This $\alpha$-threshold clusterization is only the tip of the iceberg as different near-threshold clusterings appear in $^6$He, $^6$Li, $^7$Be, $^7$Li, $^{11}$O, $^{11}$C, $^{17}$O, $^{20}$Ne, $^{26}$O, and many other nuclei. 
Various clusterings, such as $n$, $2n$, $p$, $2p$, $^2$H, $^3$He, $^3$H, $^4$He and others have been discussed. 
Near-threshold resonances play the key role for $\alpha$- and proton-capture reactions of nucleosynthesis because they  may change the astrophysical $S$-factor by orders of magnitude in the region of energies which are not accessible for the direct capture cross-section measurements.

One may ask whether the appearance of correlated many-body states close to open channels is fortuitous. It is probably obvious that they cannot result from any particular feature of the nucleon-nucleon interaction or any dynamical symmetry of the nuclear many-body problem because otherwise their appearance would be uncorrelated with the nature of different open channels. Hence, a more general arguments have to be invoked. 

It has been conjectured in SM for OQSs that the correlated (cluster) states in a vicinity of reaction channel thresholds are the generic manifestations of quantum openness of a many-body system related to the collective rearrangement of SM wave functions due to their mutual coupling via the continuum \cite{okolowicz2012,okolowicz2013}. The specific aspect in this generic phenomenon is the energetic order of particle emission  thresholds which depends on the Hamiltonian.

\begin{figure*}
\centering
\includegraphics[width=15cm,clip]{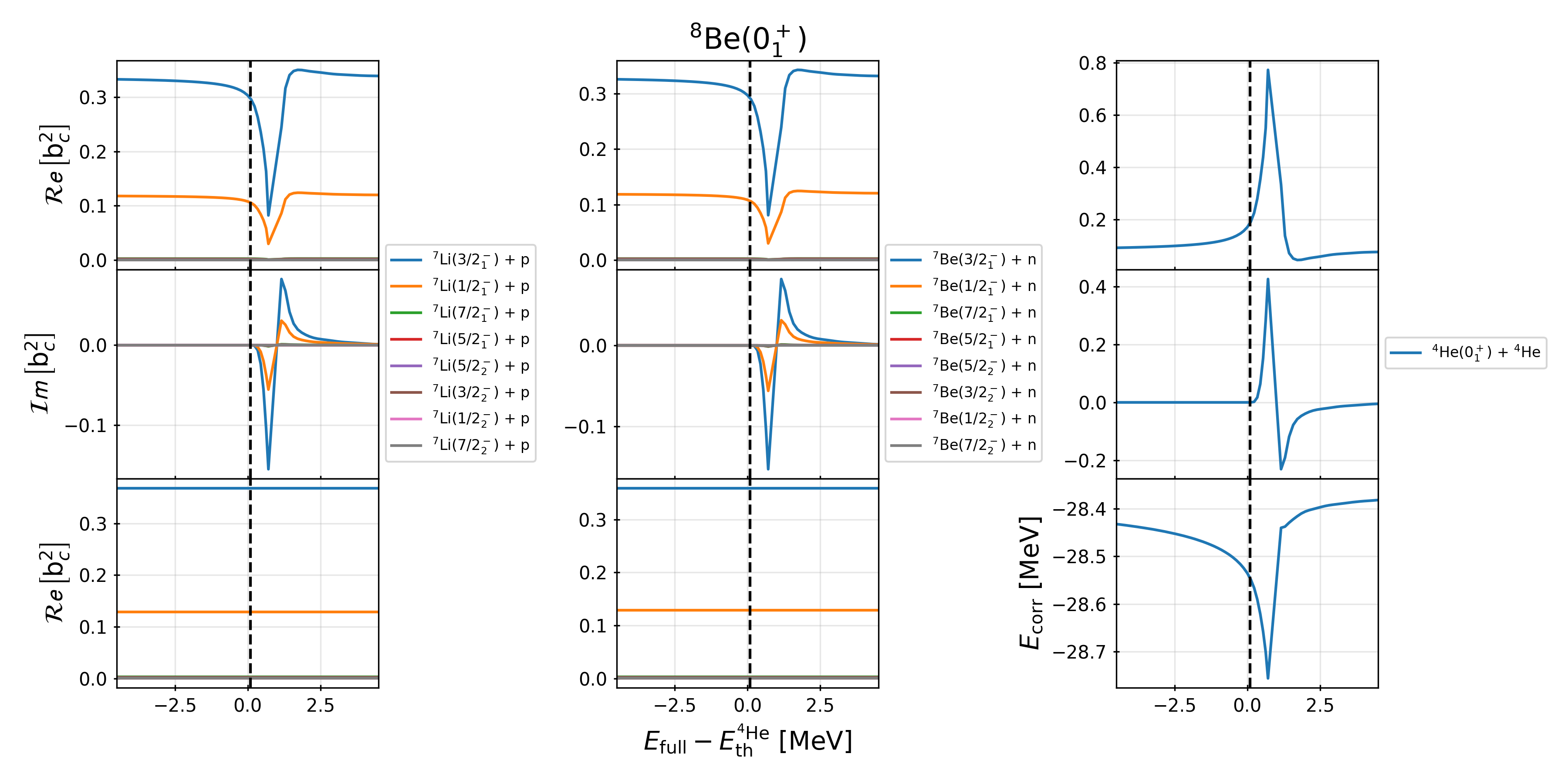}
\caption{From top to bottom: the real and imaginary parts of the channel probability weights $b^2_c$  in the $0^+_1$ state of $^8$Be. Different $^7$Li+p, $^7$Be+n, and $^4$He+$^4$He reaction channels are shown in the left, middle and right column, respectively. The meaning of different curves is given in the insert attached to each column. Each curve for proton and neutron channels represents the sum of all channels which are built on the same many-body state of $^7$Li and $^7$Be, respectively. On the bottom panels, the left and middle column show the real parts of the channel weights of the proton and neutron channels in the calculation without the channel [$^4$He($0^+_1) \otimes ^4$He($L_{\rm CM} J_{\rm int} J_{\rm P}$])]$^{0_{+}}$ , respectively, and the right column shows the continuum-coupling correlation energy. The dashed vertical line gives the GSM-CC energy of the $0^+_1$ resonance. }
\label{fig-1}      
\end{figure*}

\begin{figure*}
\centering
\includegraphics[width=15cm,clip]{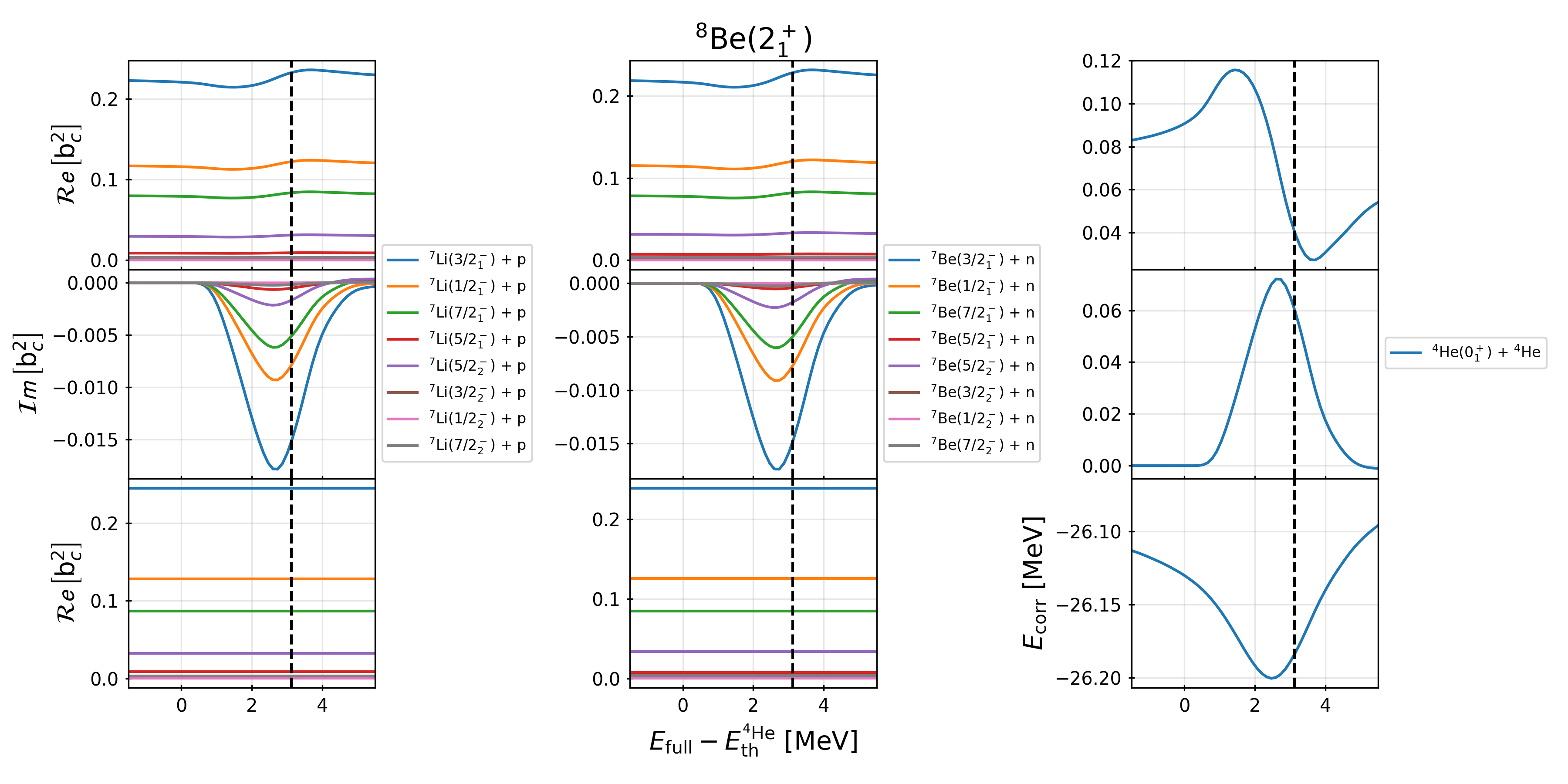}
\caption{The same as in Fig. \ref{fig-1} but for a first excited state $2_1^+$ of $^8$Be.}
\label{fig-2}       
\end{figure*}

Figure \ref{fig-1} shows the energy variation of channel weights for the ground state $0_1^+$. 
The definition of orthogonal channel probability weights is given by:
\begin{equation}
\langle \Psi|\Psi\rangle = \sum_c b^2_c = 1 + 0i
\label{eqproba}
\end{equation}
The dashed vertical line corresponds to the experimental energy of $0_1^+$. The real part of the channel weights ${\cal R}$e$[b^2_c]$ in the channel $[^4$He$(0^+_1) \otimes {^4}$He$(^1S_0)]^{0^+}$ 
has a maximum at energy E - E$^{{^4}{\rm He}}_{\rm th}$ [$^4$He$(0_1^+)] \simeq 0.7$ MeV above the $^4$He + $^4$He threshold. The maximum of the imaginary part of the channel weight ${\cal I}m[b_c^2]$ at E - E$^{{^4}{\rm He}}_{\rm th}$ [$^4$He$(0_1^+)] = 1$ MeV is shifted slightly above the maximum of the real part. 
At energies E - E$^{{^4}{\rm He}}_{\rm th}$ [$^4$He$(0_1^+)] \geq 1$ MeV, both real and imaginary parts of the $\alpha$-channel probability drop rapidly, approaching nearly a constant value. 
The maximum in the $\alpha$-channel probability is accompanied by the opposite behavior in both proton 
$[^7$Li$(3/2{^-_1}) \otimes {\rm p}(\ell j)]{^0}^+$ , $[^7$Li($1/2{^-_1} ) \otimes {\rm p}(\ell j)]^{0^+}$ 
and neutron $[^7$Be$(3/2{^-_1} ) \otimes {\rm n}(\ell j)]^{0^+}$ , $[^7$Be$(1/2{^-_1}) \otimes {\rm n}(\ell j)]^{0^+}$ channels. Other proton and neutron channels do not play any role in this phenomenon. 

It should be noticed that away from the $\alpha$-threshold, the $0^+_1$ state appears to a large extent as the SM-like state which is composed mainly by $^7$Li+p and $^7$Be+n reaction channels. Thus, the $\alpha$-clustering appears as the emergent OQS phenomenon which results from the coupling of all SM states of the same quantum numbers to the $[^4$He$(0^+_1) \otimes ^4$He$(^1S_0)]^{0^+}$ decay channel. This coupling leads to the formation of a many-body near-threshold state of the OQS which aligns with the nearby reaction channel, i.e. shares many-features of this reaction channel. This generic alignment mechanism in OQSs means that the nucleon-nucleon correlations and, hence, the structure of many-body states changes depending on the properties of the environment, i.e. nearby reaction channels and the scattering continuum. In the vicinity of the particle emission threshold, eigenstates of OQSs not only depends on the Hamiltonian but also on specific correlations that are generated in the many-body states at the proximity of the reaction channel.

In a vicinity of a threshold of a certain channel $c$, one can define the continuum-coupling correlation energy in a given many-body GSM-CC eigenstate 
$|\Psi^{J^{\pi}}\rangle $ by a difference of energies: 
\begin{equation}
E^{J^{\pi}}_{{\rm corr}, c} = \langle \Psi^{J^{\pi}}|H|\Psi^{J^{\pi}}\rangle - \langle \Psi^{J^{\pi}}_{\not c}|H|\Psi^{J^{\pi}}_{\not c} \rangle = E^{J^{\pi}}_{\rm full} - E^{J^{\pi}}_{\not c}
\label{ecorrr}
\end{equation}
where $|\Psi^{J^{\pi}}\rangle$ is the full GSM-CC solution defined as an expansion in terms of reaction channels (\ref{eq2}), and $|\Psi^{J^{\pi}}_{\not c}\rangle$ is the wave function expansion without the channel $c$. $E^{J^{\pi}}_{\rm full}$ and  $E^{J^{\pi}}_{\not c}$ are the energy eigenvalues corresponding to the eigenfunctions 
$|\Psi^{J^{\pi}}\rangle$ and $|\Psi^{J^{\pi}}_{\not c}\rangle$, respectively.

 Figure \ref{fig-2} shows the energy variation of channel probability weights for the first excited state $J^{\pi} = 2_1^+$. At low energy above the decay threshold in the channel $[^4{\rm He}(0^+_1) \otimes {^4{\rm He}}({^1D}_2)]^{2^+}$ , the $2^+_1$ state has some features of the $\alpha$-clustering. At higher energy, in an energy interval of $\sim$4 MeV, the $\alpha$-clustering is strongly reduced and the probability of the channel $[^4{\rm He}(0^+_1) \otimes {^4{\rm He}}({^1D_2})]^{2^+}$ decreases from 11$\%$ to about 5$\%$. The fall-off of the $\alpha$-clustering at higher energies is accompanied by the raise of probabilities of the one-nucleon channels built on the ground state and excited states of $^7$Be and $^7$Li. A significant decrease of the $\alpha$-cluster weight in $2^+_1$ resonance as compared to the $0^+_1$ ground state, is caused by: (i) an increased separation of the $2^+_1$ resonance from the $\alpha$-decay threshold, (ii) the centrifugal barrier, and (iii) the large width of this resonance.

\begin{figure}
\centering
\includegraphics[width=6cm,clip]{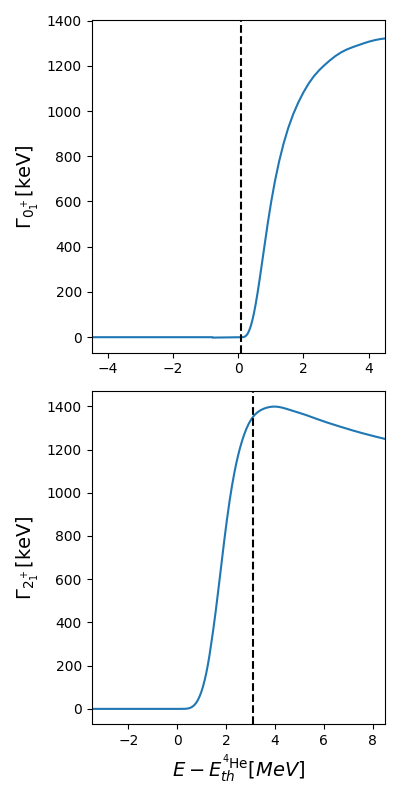}
\caption{The dependence of $\alpha$-decay width on energy distance from the $\alpha$-threshold for $0^+_1$ (left panel) and $2^+_1$ (right panel) states in $^8$Be. The vertical dashed line in both panels shows the corresponding energy of $0^+_1$ and $2^+_1$ states.}
\label{fig-3}     
\end{figure}

\vskip 0.3truecm
The energy dependence of $\alpha$-particle width of the ground state $0^+_1$ 
and the first excited state $2^+_1$ in $^8$Be is shown in Fig. \ref{fig-3}. Contrary to the $0^+_1$ state, the $\alpha$-decay width of the $2^+_1$ state  is close to its maximum at its experimental energy. One may notice that the maximum of $\alpha$-clustering in this state is at the energy where the $\alpha$-decay width is relatively small. 

\begin{figure}
\centering
\includegraphics[width=8cm,clip]{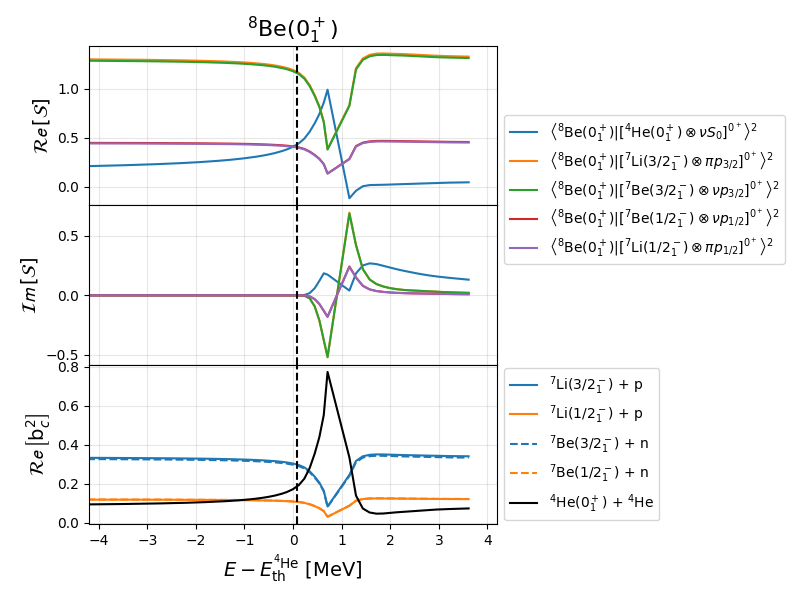}
\caption{From top to bottom: (i) the real part ${\cal R}$e$[{\cal S}^2]$ of the SFs (ii) the imaginary part ${\cal I}$m$[{\cal S}^2] $of the SFs and (iii) the real part ${\cal R}$e$[b^2_c]$ of the channel probability are given for the state $|\Psi^{0_1^+}\rangle$ of $^8$Be.}
\label{fig-1a}     
\end{figure}

\vskip 0.3truecm
Figure \ref{fig-1a} shows the dependence of spectroscopic factors (SFs) in $^8$Be on the distance to the $\alpha$-decay threshold.  The maximum of the real part of the $\alpha$-cluster SF is at $\sim$0.8 MeV what is consistent with a shift of the maximum of $b_c^2$ (see Fig. \ref{fig-1}). The behavior far away from the particle emission threshold is complicated and hence, not easy to interpret. Whereas in the vicinity of the particle emission threshold, behavior of SFs or channel probabilities $b^2_c$ is to a large extent universal, determined by the singularity of the branching point. Far away from the threshold the dependencies of those quantities are not universal; They depend on the specific features of the model and the nucleon-nucleon interactions.

\section{Conclusions}
Two generic clusterization mechanisms have been identified in atomic nucleus: the mimicry mechanism which is the quantal regime of clusterization in the low-density regime of both many-body states and reaction thresholds, and the statistical mechanism of clusterization in the high-density regime. In HI collisions at around the Fermi energy and above, it was found that the fragments are produced in the aggregation scenario \cite{PhysRevLett.86.3514}. 

\vskip 0.3truecm
 Unitarity plays an essential role in the quantum mechanism of clustering. The unitarity, which is the fundamental property of quantum mechanics, is violated in a majority of nuclear models and in that sense one can speak about the {\em unitarity crisis} in our understanding of nuclear properties.

Quantum systems in the vicinity of a particle emission threshold belong to the category of OQSs having unique properties which distinguish them from well-bound CQSs. In this respect, atomic nucleus is not an exception. In the quantal regime, the correlated (cluster) states in a vicinity of reaction channel thresholds are the generic manifestations of quantum openness.
Nuclear states in this regime not only depend on the Hamiltonian but also on the coupling to the continuum states and decay channels. This latter coupling leads to the chameleon features of resonances which have a tendency to adjust to the environment of continuum states. In all that, the unitarity plays the essential role. 

\vskip 0.3truecm
The near-threshold phenomena in atomic nuclei are unique due to the richness of nuclear interaction and the existence of nucleons in four distinct states (proton/neutron, spin-up/spin-down). In spite of large efforts, the near-threshold phenomena remain {\em terra incognita} of the nuclear physics. More efforts are necessary to the studies of (i) the collectivization of wave functions due to the coupling to decay channels, (ii) the formation of cluster/correlations, such as $^2$H, $^3$H, $^3$He, $^3$n, $^4$n, $\dots$ which carry an imprint of a nearby decay channel, (iii) the near-threshold modification of nucleon-nucleon interaction and spectroscopic factors, and (iv) the consequences of coalescing resonances in nuclear spectroscopy and reactions.

\subsection{Acknowledgments}
We wish to thank Witek Nazarewicz for useful discussions.
One of us (J.P.L.F.) wish to thank the GANIL laboratory for a hospitality when most of the results of this paper have been obtained. This work has been supported by the National Natural Science Foundation of China under Grant Nos. 11975282; the Strategic Priority Research Program of Chinese Academy of Sciences under Grant No. XDB34000000; the State Key Laboratory of Nuclear Physics and Technology, Peking University under Grant No. NPT2020KFY13.



%
%
%
\bibliographystyle{apsrev4-1}
\bibliography{refs}

%
%

\end{document}